\documentclass[12pt,draftclsnofoot,onecolumn,tbtags]{IEEEtran}

\usepackage{epsfig, graphics, color, setspace}
\usepackage{rotating, times, xspace, cite, amssymb}
\usepackage[cmex10]{amsmath}
\DeclareMathAlphabet{\mathpzc}{OT1}{pzc}{m}{it}
\DeclareMathAlphabet{\mathss}{OT1}{cmss}{m}{sl}

\renewcommand{\IEEEQED}{\IEEEQEDopen}
\newtheorem{definition}{Definition}
\newtheorem{theorem}{Theorem}
\newtheorem{lemma}[theorem]{Lemma}
\newtheorem{corollary}{Corollary}[theorem]

\newtheorem{remark}{Remark}
\allowdisplaybreaks[2]
\newcommand{\s}[1]{{\mathcal{#1}}}	
\renewcommand{\v}[1]{{\mathbf{#1}}}	
\newcommand{\vg}[1]{\boldsymbol{#1}}	

\providecommand{\Min}[1]{\min\braces{#1}}

\newcommand{\braces}[1]{{\left\{ {#1}\right\}}}
\newcommand{\paren}[1]{\left({#1}\right)}
\newcommand{\bracket}[1]{{\left [{#1}\right ]}}

\newcommand{\floor}[1]{\lfloor {#1} \rfloor}

\newcommand{\Normal}[1]{\mathcal{N}\paren{#1}}
\newcommand{\isnormal}[1]{\sim \Normal{#1}}

\newcommand{\ital}[1]{{\em #1}}
\renewcommand{\bold}[1]{{\bf #1}}
\newcommand{\tozero}{\rightarrow 0}
\newcommand{\toinf}{\rightarrow \infty}
\newcommand{\toone}{\rightarrow 1}
\newcommand{\onehalf}{\frac{1}{2}}
\newcommand{\nvar}{\sigma^2}
\newcommand{\e}{\ensuremath{\epsilon}\xspace}
\newcommand{\n}{\ensuremath{n}\xspace }
\newcommand{\Perr}[1][]{\ensuremath{P_e^{#1}}}
\newcommand{\sumton}[1][i]{\sum_{#1=1}^n}
\newcommand{\ssum}{{\textstyle \sum}}
\newcommand{\Markov}[3]{#1 \rightarrow #2 \rightarrow #3}

\newcommand{\Pmax}{\bar{P}}

\newcommand{\twon}[1]{2^{n(#1)}}
\newcommand{\ninv}[1][1]{\frac{#1}{n}}

\newcommand{\Nm}{\v{N}}

\newcommand{\Rm}{\v{R}}
\newcommand{\Wm}{\v{W}}
\newcommand{\Xm}{\v{X}}
\newcommand{\Zm}{\v{Z}}
\newcommand{\Ym}{\v{Y}}

\newcommand{\Js}{\s{J}}

\newcommand{\Ks}{\s{K}}

\newcommand{\Ws}{\s{W}}
\newcommand{\Ss}{\s{S}}
\newcommand{\Ssc}{{\Ss^c}}

\newcommand{\Gs}{\ensuremath{\s{G}}}
\newcommand{\channel}[1]{{\scriptscriptstyle \text{#1}}}
\newcommand{\Mch}{\channel{M}}
\newcommand{\Wch}{\channel{W}}

\newcommand{\CM}{C^{\Mch}}
\newcommand{\CW}{C^{\Wch}}
\newcommand{\CWs}{\tilde{C}^{\Wch}}

\newcommand{\Wmh}{\hat \Wm}

\newcommand{\Xc}{\mathfrak{X}}

\renewcommand{\Gs}[1][]{\ensuremath{\s{G}^{\scriptscriptstyle \text{#1}}}\xspace}

\newcommand{\Deltas}[1][]{\ensuremath{\Delta^{\scriptscriptstyle \text{#1}}}\xspace}
\newcommand{\NM}{\Nm_\Mch}
\newcommand{\NW}{\Nm_\Wch}
\newcommand{\NMW}{\Nm_{\Mch\Wch}}
\newcommand{\hM}{h^{\Mch}}
\newcommand{\hW}{h^{\Wch}}
\newcommand{\Xmt}{\v{\tilde X}}
\newcommand{\Nmt}{\v{\tilde N}}
\newcommand{\Ymt}{\v{\tilde Y}}
\newcommand{\Zmt}{\v{\tilde Z}}
\newcommand{\case}[5][\text{if }]{
	\begin{cases}
	#2, &#1 #3\\
	#4, &#1 #5
	\end{cases}}

\newcommand{\dssr}{$\delta$-secrecy sum-rate\xspace}
\newcommand{\dssc}{$\delta$-secrecy sum-capacity\xspace}

\newcommand{\Gsd}[1][\delta]{\ensuremath{\Gs (#1)}}
\newcommand{\GsdI}[1][\delta]{\ensuremath{\Gs[I] (#1)}}
\newcommand{\GsdC}[1][\delta]{\ensuremath{\Gs[C] (#1)}}
\newcommand{\GsdT}[1][\delta]{\ensuremath{\Gs[T] (#1)}}
\newcommand{\GsdIU}[1][\delta]{\ensuremath{\Gs[I]_\cup (#1)}}
\newcommand{\R}[1][]{\ensuremath{R^{\scriptscriptstyle \text{#1}}}\xspace}
\newcommand{\RI}[1][\delta]{\ensuremath{\R[I]_{sum} (#1)}}
\newcommand{\RC}[1][\delta]{\ensuremath{\R[C]_{sum} (#1)}}
\newcommand{\RT}[1][\delta]{\ensuremath{\R[T]_{sum} (#1)}}

\newcommand{\dinv}{\frac{1}{\delta}}
\newcommand{\ndinv}{\frac{1}{n\delta}}
\newcommand{\DeltaI}{\Deltas[I]}
\newcommand{\DeltaC}{\Deltas[C]}
\newcommand{\Deltai}{\tilde \Delta^{\scriptscriptstyle i}}
\newcommand{\Deltac}{\tilde \Delta^{\scriptscriptstyle c}}
\newcommand{\Wmi}{\Wm^{\scriptscriptstyle s}}
\newcommand{\XmS}{\Xm_\Sigma}
\newcommand{\h}{\mathss{h}}
\newcommand{\Csum}[1][\delta]{\ensuremath{C_{sum} (#1)}}
\newlength{\figsize}
\title{The Gaussian Multiple Access Wire-Tap Channel$^\dagger$
\thanks{$\dagger$ This work has been supported by NSF grant CCF-0514813 ``Multiuser Wireless Security", and was presented in part in Asilomar 2005 \cite{tekin:ASILOMAR05}, and ISIT 2006 \cite{tekin:ISIT06}.}}

\author{
\begin{minipage}[t]{1.8in}
\centering Ender~Tekin\\ \ital{tekin@psu.edu}\\
\end{minipage}%
\begin{minipage}[t]{1.8in}
\centering Aylin~Yener\\ \ital{yener@ee.psu.edu}\\ 
\end{minipage}\\
{Wireless Communications and Networking Laboratory}\\
{Electrical Engineering Department} \\
{The Pennsylvania State University}  \\
{University Park, PA 16802}\\
{\today}}

\begin{document}
\thispagestyle{headings}
\maketitle
\markboth{\ital{Submitted to IEEE Transactions on Information Theory May 2006, Revised March 2008.}}{}

\vspace{-.5in}
\begin{abstract}
We consider the Gaussian Multiple Access Wire-Tap Channel (GMAC-WT).  In this scenario,
multiple users communicate with an intended receiver in the presence of an intelligent and informed wire-tapper who receives a degraded version of the signal at the receiver. We define suitable security measures for this multi-access environment. Using codebooks generated randomly according to a Gaussian distribution, achievable secrecy rate regions are identified using superposition coding and TDMA coding schemes. An upper bound for the secrecy sum-rate is derived, and our coding schemes are shown to achieve the sum capacity.  Numerical results showing the new rate region are presented and compared with the capacity region of the Gaussian Multiple-Access Channel (GMAC) with no secrecy constraints, quantifying the price paid for secrecy.
\end{abstract}
\section{Introduction}

Shannon, in \cite{shannon:secrecy}, analyzed secrecy systems in communications and showed that, to achieve perfect secrecy of communications, the conditional probability of the \ital{cryptogram given a message} must be independent of the actual transmitted message. In \cite{wyner:wiretap}, Wyner applied this concept to the discrete memoryless channel, with a wire-tapper who has access to a degraded version of the intended receiver's signal.  He measured the amount of ``secrecy" using the conditional entropy $\Delta$, the conditional entropy of the transmitted message given the received signal at the wire-tapper.  The region of all possible rate/equivocation, $(R,\Delta)$, pairs is determined, and the existence of a positive \ital{secrecy capacity}, $C_s$, for communication below which it is possible to limit the rate of information leaked to the wire-tapper to arbitrarily small values, is shown \cite{wyner:wiretap}.  Carleial and Hellman, in \cite{hellman-carleial:wiretap}, showed that it is possible to send several low-rate messages, each completely protected from the wire-tapper individually, and use the channel at close to capacity.  The drawback is, in this case, if any of the messages are revealed to the wire-tapper, the others might also be compromised.  In \cite{leung-hellman:gaussianwiretap}, the authors extended Wyner's results to Gaussian channels and also showed that Carleial and Hellman's results in \cite{hellman-carleial:wiretap} also held for the Gaussian channel. Csisz\'ar and K\"orner, in \cite{csiszar-korner:confbroadcast}, showed that Wyner's results can be extended to weaker, so called ``less noisy" and ``more capable" channels. Furthermore, they analyzed the more general case of sending common information to both the receiver and the wire-tapper, and private information to the receiver only.

More recently, the closely related problem of common randomness and secret key generation has gathered attention.  Maurer, \cite{maurer:secretkeypublicdiscussion}, and Bennett et. al., \cite{bennettetal:genprivacy}, have focused on the process of ``distilling" a secret key between two parties in the presence of a wire-tapper.  In this scenario, the wire-tapper has partial information about a common random variable shared by the two parties, and the parties use their knowledge of the wire-tapper's limitations to generate a secret key.  Reference \cite{bennettetal:genprivacy} breaks this down into three main steps: (i) advantage distillation: where the two parties have zero wiretap capacity and need to find some way of creating an advantage over the wire-tapper, (ii) information reconciliation: where the secret key decided by one of the partners is communicated to the other partner and the wire-tapper is still left with only partial information about it, (iii) privacy amplification: where a new secret key is generated from the previous one about which the wire-tapper has negligible information.  In \cite{maurer:secretkeypublicdiscussion}, it was shown that for the case when the wire-tap channel capacity is zero between two users, the existence of a ``public" feedback channel that the wire-tapper can also observe can nevertheless enable the two parties to be able to generate a secret key with perfect secrecy.  This discussion was then furthered by \cite{ahlswede-csiszar:CR1} and \cite{ahlswede-csiszar:CR2} where the secrecy key capacities and \ital{common randomness} capacities, the maximum rates of common randomness that can be generated by two terminals, were developed for several models.  It was also argued in \cite{maurer-wolf:weaktostrongsecrecy}, that the secrecy constraint developed by Wyner and later utilized by Csisz\'ar and K\"orner was ``weak" since it only constrained the rate of information leaked to the wire-tapper, rather than the total information.  It was shown that Wyner's scenario could be extended to ``strong" secrecy for discrete channels with no loss in achievable rates, where the secrecy constraint is placed on the total information obtained by the wire-tapper, as the information of interest might be in the small amount leaked.  This corresponds to making the leaked information go to zero exponentially rather than just inversely with $n$.  Maurer then examined the case of active adversaries, where the wire-tapper has read/write access to the channel in \cite{maurer-wolf:SecretKey1}-- \nocite{maurer-wolf:SecretKey2}\cite{maurer-wolf:SecretKey3}.  Venkatesan and Anantharam examined the cases where the two terminals generating common randomness were connected by different DMC's in \cite{venkatesan-anantharam:CRcap-pair} and later generalized this to a network of DMC's connecting any finite number of terminals in \cite{venkatesan-anantharam:CRcap-network}. Csisz\'ar and Narayan extended Ahslwede and Csisz\'ar's previous work to multiple-terminals by looking at what a helper terminal can contribute in \cite{csiszar-narayan:CRhelper}, and the case of multiple terminals where an arbitrary number of terminals are trying to distill a secret key and a subset of these terminals can act as helper terminals to the rest in \cite{csiszar-narayan:secrecycap-multi}.

More recently, the notion of the wire-tap channel has been extended to parallel channels and fading channels \cite{yamamoto:secretsharing, yamamoto:secretsharinggaussian, barros:fadingwiretap, gopala:fadingsecISIT, liang:Allerton06a, zang:parallelsecrecy}. Broadcast and interference channels with confidential messages were considered in \cite{liuetal:IBCconf}.  References \cite{liang:genMACconfPAP, liuetal:MACconf} examined the so called multiple access channel with confidential messages, where two transmitters try to keep their messages secret from each other while communicating with a common receiver, finding an achievable region, and the capacity region for some special cases.

In this paper, we define the Gaussian Multiple Access Wire-Tap Channel (GMAC-WT) where multiple users are transmitting to a base station in the presence of AWGN, and a wiretapper receives a noisy version of the signal received at the base station.  For this new multi-transmitter secrecy paradigm, we first define two separate secrecy constraints, which we call the \ital{individual} and \ital{collective} secrecy constraints.  These are (i) the normalized entropy of any set of messages conditioned on the transmitted codewords of the other users and the received signal at the wire-tapper, and (ii) the normalized entropy of any set of messages conditioned on the wire-tapper's received signal.  The first set of constraints is more conservative to ensure secrecy of any subset of users even when the remaining users are compromised.  The second set of constraints ensures the collective secrecy of any set of users, utilizing the secrecy of the remaining users.  In \cite{tekin:ASILOMAR05}, we concerned ourselves mainly with the \ital{perfect secrecy rate region} for both sets of constraints.  In this paper, we consider the general case where a pre-determined level of secrecy is provided.  Using codebooks generated according to a Gaussian distribution and superposition coding, we find the achievable \ital{secrecy rate regions} for each constraint, where users can communicate with arbitrarily small probability of error with the intended receiver, while the wire-tapper is kept ignorant to a pre-determined level.  This scheme achieves the secrecy sum capacity for collective constraints.  We also find a secrecy rate region using TDMA and the results of \cite{leung-hellman:gaussianwiretap} for a single-user.  This scheme achieves secrecy sum capacity for both constraints, but is smaller than the region for collective constraints. When individual constraints are considered, the achievable region is the convex hull of the union of the superposition coding and TDMA regions.

Finally, a word on notation: Throughout the paper, we denote vectors with bold letters, as well as letters with superscripts as the length of the vector whenever necessary.  Sets are denoted using a script font.  Also, for random variables $X,Y$, we use $H(X)$ to refer to the entropy of $X$ when it is discrete, $\h(X)$ to refer to the differential entropy when $X$ is continuous, and $I(X;Y)$ to refer to the mutual information of $X,Y$.  All the logarithms are taken to base 2, such that the resulting information theoretic quantities are in bits.
\section{System Model and Problem Statement}
\label{sec:system}
\begin{figure}[p]
\centering
\includegraphics[width=\figsize,angle=0]{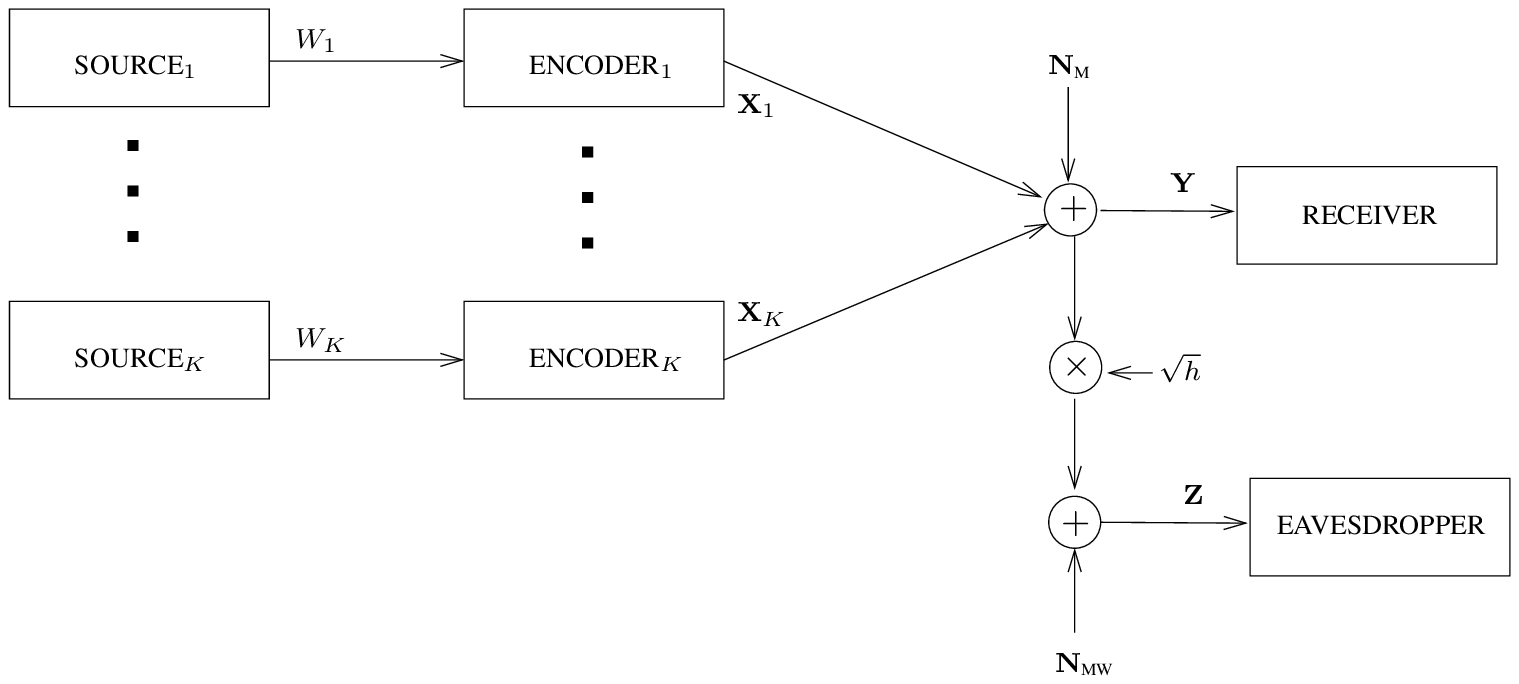}
\caption{Equivalent GMAC-WT System Model.}
\label{fig:gmacwtdeg}
\end{figure}

We consider $K$ users communicating with a receiver in the presence of a wire-tapper in a Gaussian channel.  The model under consideration is presented in Figure \ref{fig:gmacwtdeg}.  In general, transmitter $k \in \Ks \triangleq \{1,\dotsc,K\}$ chooses a message $W_k$ from a set of equally likely messages $\Ws_k=\{1, \dotsc, M_k\}$. The messages are encoded using $(2^{nR_k},n)$ codes into $\{\tilde X_k^n(W_j)\}$, where $R_k=\ninv \log_2 M_k$. The encoded messages $\{\Xmt_k\}=\{\tilde X_k^n\}$ are then transmitted, and the intended receiver and the wire-tapper get $\Ymt=\tilde Y^n$ and $\Zmt=\tilde Z^n$, respectively.  The receiver decodes $\Ymt$ to get an estimate of the transmitted messages, $\Wmh$.  We would like to communicate with the receiver with arbitrarily low probability of error, while maintaining secrecy to a pre-determined level $\delta$.  The signals at the intended receiver and the wiretapper are then given by
\begin{subequations}
\label{eqn:YZ}
\begin{align}
\Ymt &= \ssum_{k=1}^K \sqrt{\hM_k} \Xmt_k + \Nmt_\Mch \\
\Zmt &= \ssum_{k=1}^K \sqrt{\hW_k} \Xmt_k + \Nmt_\Wch
\end{align}
\end{subequations}
where $\Nmt_\Mch,\Nmt_\Wch$ are the AWGN at the intended receiver and eavesdropper, respectively.  Each component of $\Nmt_\Mch \isnormal{0,\nvar_\Mch}$ and $\Nmt_\Wch \isnormal{0,\nvar_\Wch}$.  The transmit power constraints are given by $\ninv \sumton{\tilde X_{ki}^2} \le \tilde \Pmax_k, \; k=1,\dotsc,K$.  Faithful to Wyner's terminology, we shall call this channel model, The Gaussian Multiple-Access Wire-Tap Channel (GMAC-WT).  

Similar to the scaling transformation to put an interference channel in standard form, \cite{carleial:interference}, we can represent any GMAC-WT by an equivalent standard form, \cite{tekin:ISIT06}:
\begin{subequations}
\label{eqn:YZstd}
\begin{align}
\Ym &= \ssum_{k=1}^K \Xm_k + \NM \\
\Zm &= \ssum_{k=1}^K \sqrt{h_k} \Xm_k + \NW
\end{align}
\end{subequations}
where
\begin{itemize}
\item the codewords $\{\tilde \Xm\}$ are scaled to get $\Xm_k = \sqrt{\frac{\hM_k}{\nvar_\Mch}}\tilde \Xm_k$;
\item the new power constraints are $\Pmax_k = \frac{\hM_k}{\nvar_\Mch}\tilde \Pmax_k$;
\item the new wiretapper channel gains are $h_k = \frac{\hW_k \nvar_\Mch}{\hM_k \nvar_\Wch}$;
\item the AWGN are normalized by $\NM = \frac{1}{\nvar_\Mch} \Nmt_\Mch$ and $\NW = \frac{1}{\nvar_\Wch}\Nmt_\Wch$.
\end{itemize}

In this paper, we shall examine the case where the wire-tapper receives a \ital{stochastically degraded} version of the signal received at the legitimate receiver, i.e., there exists a distribution $\tilde p(z|y)$ such that we can write $p(z|x_1,\dotsc,x_K)=\int_{-\infty}^{\infty} p(y|x_1,\dotsc,x_K)\tilde p(z|y) dy$. Similar to the broadcast channel, since the legitimate receiver and the eavesdropper do not cooperate, noise correlations do not play a role, and as a result, the capacity of the stochastically degraded wire-tap channel is the same as that of the \ital{physically degraded} wire-tap channel, which means $p(y,z|x_1,\dotsc,x_K) =p(y|x_1,\dotsc,x_K)p(z|y)$.  It can easily be shown that the wire-tapper gets a stochastically degraded version of the receiver's signal if $h_1=\dotsc=h_K \equiv h < 1$.  Equivalently, we consider the physically degraded case, where the wire-tapper's received signal is a noisier version of the legitimate receiver's scaled received signal, $\Zm=\sqrt{h} \Ym + \NMW$, where $\NMW$ has each component $\isnormal{0,1-h}$ and is independent of $\Ym$.  This model is illustrated in Figure \ref{fig:gmacwtdeg}.  In practical situations, we can think of this as the wire-tapper being outside of a controlled indoor environment, such as in \cite{yakovlev:wiretapradio} or just being able to wire-tap the receiver rather than receive the signals itself.

\subsection{The Secrecy Measures}
Letting $\Delta_\Ss$ be our secrecy constraint for any subset $\Ss$ of users, we require that $\Delta_\Ss \ge \delta$ for all sets $\Ss \subseteq \Ks$, with $\delta \in [0,1]$ as the required level of secrecy.  $\delta=1$ corresponds to \ital{perfect secrecy}, where the wire-tapper is not allowed to get any information; and $\delta=0$ corresponds to no secrecy constraint. To that end, we define two sets of secrecy constraints using the normalized equivocations for sets of users.  These are: 

\subsubsection{Individual Secrecy}
We define the individual secrecy measure for a subset of users, $\Ss \subseteq \Ks$, as
\begin{equation}
\label{eqn:defDeltaI}
\DeltaI_\Ss \triangleq \frac{H(\Wm_\Ss|\Xm_\Ssc,\Zm)}{H(\Wm_\Ss)}
\quad \forall \Ss \subseteq \Ks=\{1,\dotsc,K\}
\end{equation}
where $\Wm_\Ss = \{W_k\}_{k \in \Ss}$.  $\DeltaI_\Ss$ denotes the normalized entropy of the transmitted messages of a set $\Ss$ of users, given the received signal at the wire-tapper as well as the remaining users' transmitted symbols. As our secrecy criterion, this guarantees that the rate of information leaked to the wire-tapper from a group of users is limited even if all the other users' transmitted codewords are compromised. Note that this is a stronger constraint than $H(\Wm_\Ss|\Wm_\Ssc,\Zm)$, as $H(\Wm_\Ss|\Wm_\Ssc,\Zm) \ge H(\Wm_\Ss|\Wm_\Ssc,\Xm_\Ssc,\Zm)=H(\Wm_\Ss|\Xm_\Ssc,\Zm)$. In addition, from a practical point of view, if the transmitted messages are compromised either due to byzantine users or some other side information allowing the eavesdropper to decode the transmitted messages of a group of users, there is no reason to expect that the transmitted codewords are not known to the eavesdropper. Thus, this represents a scenario where users do not have to trust each other.

We note that if the individual secrecy constraints for all users in the set $\Ss$ are satisfied, i.e., $\Delta_k \ge \delta, \, \forall k \in \Ss$, then the constraint for set $\Ss$ is also satisfied.  To see this, without loss of generality, let $\Ss=1,\dotsc,S$ where $S \le K$ and assume $\frac{H(W_k|\Xm_{k^c},\Zm)}{H(W_k)} \ge \delta$. We can write
\begin{align}
H(\Wm_\Ss|\Xm_{\Ssc},\Zm) &= \sum_{k=1}^S H(W_k|W^{k-1}, \Xm_{\Ssc}, \Zm) \\
	&\label{eqn:indsec1} \ge \sum_{k=1}^S H(W_k|W^{k-1}, \Xm_{k^c},\Zm)\\
	&\label{eqn:indsec2} =\sum_{k=1}^S H(W_k|\Xm_{k^c},\Zm)\\
	&\label{eqn:indsec3} \ge \sum_{k=1}^S \delta H(W_k) \\
	&\label{eqn:indsec}=\delta H(\Wm_\Ss)
\end{align}
where \eqref{eqn:indsec1} follows using conditioning, \eqref{eqn:indsec2} is due to the fact that $W_j$ is conditionally independent of all $W_k$ given $\Xm_k,\Zm$.  \eqref{eqn:indsec3} comes from our assumption that for all $k \in \Ss$, $\DeltaI_k \ge \delta$. Thus, for any subset of users the individual secrecy constraints for all users also guarantee the joint secrecy of the same level for the entire set.

\subsubsection{Collective Secrecy}
Clearly \eqref{eqn:defDeltaI} is a conservative measure where users do not place any trust on each other.  We now define a revised secrecy measure to take into account the multi-access nature of the channel where users rely on others to achieve secrecy for the whole group.
\begin{equation}
\label{eqn:defDeltaC}
\DeltaC_\Ss \triangleq \frac{H(\Wm_\Ss|\Zm)}{H(\Wm_\Ss)} \quad \forall \Ss \subseteq \Ks
\end{equation}

Using this constraint guarantees that each subset of users maintains a level of secrecy greater than $\delta$.  Since this must be true for all sets of users, collectively the system has at least the same level of secrecy.  However, if a group of users are somehow compromised, the remaining users may also be vulnerable.  We require the secrecy constraint to be satisfied separately for each $\Ss \subseteq \Ks$, since otherwise it is possible to have $\DeltaC_\Ss \ge \delta$, but $\DeltaC_{\Js} < \delta$ for some $\Js \subset \Ss$.  However, if $\delta=1$, i.e. when we require perfect secrecy, we can show that $\DeltaC_\Ks \ge 1-\e \Rightarrow \DeltaC_\Ss\ge 1-\e'$ for all $\Ss \subseteq \Ks$, where $\e' \tozero$ as $\e \tozero$.  To see this, write
\begin{align}
H(\Wm_\Ss|\Zm) + H(\Wm_\Ssc) 
	&\ge H(\Wm_\Ss|\Zm) + H(\Wm_\Ssc|\Wm_\Ss,\Zm) \label{eqn:colsecder0}\\
	&=H(\Wm_\Ks|\Zm) \label{eqn:colsecder1}\\
	&\ge(1-\e) H(\Wm_\Ks) \label{eqn:colsecder2}\\
	&=(1-\e) H(\Wm_\Ss)+(1-\e) H(\Wm_\Ssc) \label{eqn:colsecder3}
\end{align}
where \eqref{eqn:colsecder1} follows from the chain rule of entropy and \eqref{eqn:colsecder2} from the requirement for perfect secrecy.  Comparing the left hand side of \eqref{eqn:colsecder0} and \eqref{eqn:colsecder3}, since conditioning cannot increase entropy, we see that we have to have
\begin{equation}
\label{eqn:colsec}
\frac{H(\Wm_\Ss|\Zm)}{H(\Wm_\Ss)} \ge 1-\e'
\end{equation}
where $\e' \triangleq \paren{1+\frac{H(\Wm_\Ssc)}{H(\Wm_\Ss)}}\e\tozero$ as $\e\tozero$. Thus, perfect secrecy for the ensemble of users guarantees perfect secrecy for all subsets of users.

\subsection{The $\delta$-secret Achievable Rate Region}
\begin{definition}[Achievable Rates with $\delta$-secrecy]
\label{def:achrate}
Let $\xi=I$ if using individual constraints, and $\xi=C$ if using collective constraints. The rate $K$-tuple $\Rm=(R_1,\dotsc,R_K)$ is said to be \ital{achievable with $\delta$-secrecy} under constraint $\xi$, if for any given $\e>0$ there exists a code of sufficient length \n such that
\begin{align}
\ninv \log_2 M_k &\ge R_k - \e, \qquad k=1,\dotsc,K\\
\Perr &\le \e \\
\Delta_\Ss^\xi &\ge \delta, \qquad \forall \Ss \subseteq \Ks
\end{align}where user $k$ chooses one of $M_k$ symbols to transmit according to the uniform distribution, $\Delta_\Ss^\xi$ denotes the secrecy constraint, and is given by \eqref{eqn:defDeltaI} if $\xi=I$, and by \eqref{eqn:defDeltaC} if $\xi=C$.  We will call the set of all achievable rates with $\delta$-secrecy, the \ital{$\delta$-secret achievable rate region}, and denote it $\s{C}^{\xi}(\delta)$.
\end{definition}

\subsection{Some Preliminary Definitions}
Before we state our results, we also define the following quantities for any $\Ss \subseteq \Ks$.
\begin{align}
P_\Ss &\triangleq \ssum_{k \in \Ss} P_k \\
R_\Ss &\triangleq \ssum_{k \in \Ss} R_k \\
\CM_\Ss &\triangleq \onehalf \log \paren{1+P_\Ss} \\
\CW_\Ss &\triangleq \onehalf \log \paren{1+h P_\Ss} \\
\CWs_\Ss &\triangleq \onehalf \log \paren{1+\frac{h P_\Ss}{1+h P_\Ssc}}
\end{align}

Alternately, we also use the subscript $sum$ when $\Ss=\Ks$.
\section{Achievable $\delta$-Secrecy Rate Regions}
\label{sec:ach}
In this section, we find a set of achievable rates using Gaussian codebooks, which we denote by \Gsd.  We first give the achievable regions satisfying the individual and collective secrecy constraints, denoted by $\GsdI$ and $\GsdC$ respectively, using superposition coding. We then give the region when TDMA is used, denoted $\GsdT$, and satisfies both secrecy constraints.  For the collective secrecy constraints, the TDMA region is seen to be smaller than the superposition coding region.  For the individual constraints, the achievable region is the convex hull of the union of the superposition and TDMA regions, denoted $\GsdIU$.

\subsection{Individual Secrecy}
In \cite{leung-hellman:gaussianwiretap}, it has been shown that Gaussian codebooks can be used to maintain secrecy for the single user Gaussian wire-tap channel. Using a similar approach, we present an achievable region for $\delta$-secrecy using individual secrecy constraints in Theorem \ref{thm:GI}.
\begin{theorem}
\label{thm:GI}
The following region is achievable with $\delta$-secrecy for the GMAC-WT using Gaussian codebooks.
\begin{equation}
\label{eqn:GI}
\GsdI = \braces{\Rm \colon R_\Ss \le 
	\Min{\CM_\Ss, \dinv \paren{\CM_\Ss - \sum_{k \in \Ss} \CW_k}} \quad \forall \Ss \subseteq \Ks}
\end{equation}
\end{theorem}
\begin{IEEEproof}
\bold{Coding Scheme:} Let $\Rm=(R_1,\dotsc,R_K)$ satisfy \eqref{eqn:GI}.  For each user $k \in \Ks$, consider the scheme:
\begin{IEEEenumerate}
\item Let $M_k=\twon{R_k-\e'}$ where $0 \le \e' < \e$.  Let $M_k=M_{ks}M_{k0}$ where, for some $1 \ge \mu_k \ge \delta$ to be chosen later, $M_{ks}=M_k^{\mu_k}, \, M_{k0}=M_k^{1-\mu_k}$. We then have $R_k=R_{ks}+R_{k0}+\e'$, where	$R_{ks}=\ninv \log M_{ks}$ and $R_{k0}=\ninv \log M_{k0}$.  We can choose $\e'$ and $n$ to ensure that $M_{ks},M_{k0}$ are integers.

\item	Generate $3$ codebooks $\Xc_{ks},\Xc_{k0}$ and $\Xc_{kx}$.  $\Xc_{ks}$ consists of $M_{ks}$	codewords, each component of which is drawn from $\Normal{0,\lambda_{ks}P_k -\varepsilon}$. Codebook $\Xc_{k0}$ has $M_{k0}$ codewords with each component drawn from $\Normal{0,\lambda_{k0} P_k-\varepsilon}$ and $\Xc_{kx}$ has $M_{kx}$ codewords with each component drawn from $\Normal{0,\lambda_{kx} P_k-\varepsilon}$. Here, $\varepsilon$ is an arbitrarily small number to ensure that the power constraints are satisfied with high probability, and $\lambda_{ks}{+}\lambda_{k0}{+}\lambda_{kx}{=}1$. Define $R_{kx}=\ninv \log M_{kx}$ and $M_{kt}=M_k M_{kx}$.

\item Each message $W_k \in \{1,\dotsc,M_k\}$ is mapped into a message vector	$\Wm_k=(W_{ks},W_{k0})$ where $W_{ks}\in \{1,\dotsc,M_{ks}\}$ and $W_{k0}\in \{1,\dotsc,M_{k0}\}$.	Since $W_k$ is uniformly chosen, $W_{ks},W_{k0}$ are also uniformly distributed.

\item To transmit message $W_k \in \{1,\dotsc,M_k\}$, user $k$ finds the $2$ codewords 
corresponding to components of $\Wm_k$ from $\Xc_{ks}$ and $\Xc_{k0}$, and also uniformly chooses a codeword from $\Xc_{kx}$. He then adds all these codewords and transmits the resulting codeword, $\Xm_k$, so that we are actually transmitting one of $M_{kt}$ codewords. Let $R_{kt} = \ninv \log M_{kt} +\e'= R_{ks}+R_{k0}+R_{kx}+\e'$. 
\end{IEEEenumerate}

Specifically, the rates are chosen to satisfy $\forall \Ss \subseteq \Ks$:
\begin{align}
\label{eqn:achIRs} &\ssum_{k \in \Ss} R_{ks} = \ssum_{k \in \Ss} \mu_k R_k 
	\le \CM_\Ss-\sum_{k \in \Ss} \CW_k\\
\label{eqn:achIR0} &R_{k0}+R_{kx} 
	= (1-\mu_k)R_k+R_{kx} = \CW_k, \quad \forall k \in \Ss \\
\label{eqn:achIRt} &\ssum_{k \in \Ss} R_{kt} 
	= \ssum_{k \in \Ss} \bracket{R_k+R_{kx}} \le \CM_\Ss
\end{align}
Consider the sub-code $\{\Xc_{ks}\}_{k=1}^K$.  From this point of view, the coding scheme described is equivalent to each user $k \in \Ks$ selecting one of $M_{ks}$ messages, and sending a uniformly chosen codeword from among $M_{k0}M_{kx}$ codewords for each.  Let  $\Deltai_k=\frac{H(W_{ks}|\Xm_{k^c},\Zm)}{H(W_{ks})}$ and write:
\begin{align}
H(W_{ks}|\Xm_{k^c},\Zm) &= H(W_{ks}|\Xm_{k^c}) - I(W_{ks};\Zm|\Xm_{k^c}) \\
	&= H(W_{ks}) - I(W_{ks};\Zm|\Xm_{k^c}) \label{eqn:achprf0i}\\
	&= H(W_{ks}) - I(W_{ks};\Zm|\Xm_{k^c}) + I(W_{ks};\Zm|\Xm_\Ks) \label{eqn:achprf0ii}\\
	&= H(W_{ks}) - \h(\Zm|\Xm_{k^c}) {+} \h(\Zm|W_{ks}{,}\Xm_{k^c}) 
		{+} \h(\Zm|\Xm_\Ks) {-} \h(\Zm|\Xm_\Ks{,}W_{ks})\\
   &=H(W_{ks}) - I(\Xm_k;\Zm|\Xm_{k^c}) + I(\Xm_k;\Zm|W_{ks},\Xm_{k^c})
\end{align}
where \eqref{eqn:achprf0i} follows from the fact that the secret message of user $k$ is independent of the codewords of the remaining users, and \eqref{eqn:achprf0ii} follows since the received signal at the eavesdropper is independent of the transmitted secret messages given the actual transmitted codewords.  Thus, we have:
\begin{equation}
\Deltai_k = \frac{H(W_{ks}|\Xm_{k^c},\Zm)}{H(W_{ks})} 
	= 1-\frac{I(\Xm_k;\Zm|\Xm_{k^c}) - I(\Xm_k;\Zm|W_{ks},\Xm_{k^c})}{H(W_{ks})}
	\label{eqn:achprf1i}
\end{equation}

By the converse to the coding theorem for the Gaussian Multiple Access Channel, we have $I(\Xm_k;\Zm|\Xm_{k^c}) \le n\CW_k$. We can also write
\begin{equation}
I(\Xm_k;\Zm|W_{ks},\Xm_{k^c})=H(\Xm_k|W_{ks},\Xm_{k^c})-H(\Xm_k|W_{ks},\Xm_{k^c},\Zm)
\end{equation}

For each secret message of user $k$, our coding scheme implies that it sends one of $M_{k0}M_{kx}$ possible codewords.  By choosing $R_{k0},R_{kx}$ to satisfy \eqref{eqn:achIR0}, we guarantee that
\begin{equation}
H(\Xm_k|W_{ks},\Xm_{k^c}) = H(\Xm_k|W_{ks}) = n\CW_k
\end{equation}

Also,
\begin{equation}
H(\Xm_k|W_{ks},\Xm_{k^c},\Zm) \le n\delta_n
\end{equation}
where $\delta_n \tozero$ due to Fano's Inequality. This stems from the fact that given $W_{ks}$, the sub-code for user $k$ is, with high probability, a ``good" code for the wiretapper.  Combining these in \eqref{eqn:achprf1i}, we can write
\begin{equation}
\label{eqn:achprf2i}
\Deltai_k \ge 1-\frac{n \CW_k - n \CW_k + n\delta_n}{H(W_{ks})} =1-\e
\end{equation}
where $\e = \frac{\delta_n}{R_{ks}} \tozero$ as $n \toinf$.

Then, we can write
\begin{equation}
\label{eqn:achprf3i}
\DeltaI_k = \frac{H(W_k|\Xm_{k^c},\Zm)}{H(W_k)} 
	\ge \frac{H(W_{ks}|\Xm_{k^c},\Zm)}{H(W_k)} 
	\ge \frac{(1-\e)H(W_{ks})}{H(W_k)}
	\ge \frac{(1-\e) \mu_k R_k}{R_k}
	\ge \delta
\end{equation}

Since \eqref{eqn:achprf3i} holds for all $k=1,\dotsc,K$, from \eqref{eqn:indsec} we have
$\DeltaI_\Ss \ge \delta, \; \forall \Ss \subseteq \Ks$.
\end{IEEEproof}
\begin{remark}
In this case, the maximum \dssr achievable is given by
\begin{equation}
\label{eqn:RI}
\RI = \Min{\CM_\Ks,\dinv \bracket{\CM_\Ks - \sum_{k=1}^K \CW_k}}
\end{equation}
Observe that there is a reduction of $\sum_{k=1}^K \CW_k \ge \CW_\Ks$ in the \dssr due to secrecy constraints when the second term is the minimum. Also observe that the transmission of all the users with their maximum power may not be optimal for this case. 
\end{remark}

\subsection{Collective Secrecy}
\begin{theorem}
\label{thm:GC}
We can transmit with $\delta$-secrecy using Gaussian codebooks at rates in the region $\GsdC$ defined as 
\begin{equation}
\label{eqn:GC}
\GsdC \triangleq \braces{\Rm \colon R_\Ss \le \Min{\CM_\Ss, 
\dinv \bracket{\CM_\Ss - \CWs_\Ss}} \quad \forall \Ss \subseteq \Ks}
\end{equation}
\end{theorem}
\begin{IEEEproof}
Let $\Rm=(R_1,\dotsc,R_K)$ satisfy \eqref{eqn:GC} and assume the coding scheme is the same as described in the individual constraints case, except that instead of \eqref{eqn:achIRs}--\eqref{eqn:achIRt}, we will choose the rates such that for all $\Ss \subseteq \Ks$,
\begin{align}
\label{eqn:achRs} &\ssum_{k \in \Ss} R_{ks} = \ssum_{k \in \Ss} \mu_k R_k 
	\le \CM_\Ss-\CWs_\Ss \\
\label{eqn:achR0} &\ssum_{k=1}^K \bracket{R_{k0}+R_{kx}} 
	= \ssum_{k=1}^K \bracket{(1-\mu_k)R_k+R_{kx}} = \CW_\Ks \\
\label{eqn:achRt} &\ssum_{k \in \Ss} R_{kt} 
	= \ssum_{k \in \Ss} \bracket{R_k+R_{kx}} \le \CM_\Ss
\end{align}
From \eqref{eqn:achRt} and the GMAC coding theorem, with high probability the receiver can decode the codewords with low probability of error. To show $\DeltaC_\Ss \ge \delta, \; \forall \Ss \subseteq \Ks$, we concern ourselves only with MAC sub-code $\{\Xc_{ks}\}_{k=1}^K$.  From this point of view, the coding scheme described is equivalent to each user $k \in \Ks$ selecting one of $M_{ks}$ messages, and sending a uniformly chosen codeword from among $M_{k0}M_{kx}$ codewords for each.  Let $\Wmi_\Ss = \{W_{ks}\}_{k \in \Ss}$ and $\Deltac_\Ss=\frac{H(\Wmi_\Ss|\Zm)}{H(\Wmi_\Ss)}$ and define $\XmS=\sum_{k=1}^K \Xm_k$. For $\Ks$ write
\begin{align}
H(\Wmi_\Ks|\Zm) &= H(\Wmi_\Ks) - I(\Wmi_\Ks;\Zm) \\
	&= H(\Wmi_\Ks) - I(\Wmi_\Ks;\Zm) + I(\Wmi_\Ks;\Zm|\XmS) \label{eqn:achprf0c} \\
	&= H(\Wmi_\Ks) - \h(\Zm) + \h(\Zm|\Wmi_\Ks) + \h(\Zm|\XmS) - \h(\Zm|\Wmi_\Ks,\XmS) \\
	&= H(\Wmi_\Ks) - I(\XmS;\Zm) + I(\XmS;\Zm|\Wmi_\Ks)
\end{align}
where \eqref{eqn:achprf0c} follows from $\Markov{\Wmi_\Ks}{\XmS}{\Zm}$.  Therefore, we have:
\begin{equation}
\Deltac_\Ks	=\frac{H(\Wmi_\Ks|\Zm)}{H(\Wmi_\Ks)}
	= 1- \frac{I(\XmS;\Zm)-I(\XmS;\Zm|\Wmi_\Ks)}{H(\Wmi_\Ks)} \label{eqn:achprf1}
\end{equation}

Consider the two terms individually.  First, we have the sum-rate bound of the multiple access channel to the eavesdropper:
\begin{equation}
\label{eqn:achprf3}
I(\XmS;\Zm) \le n\CW_\Ks
\end{equation}

$I(\XmS;\Zm|\Wmi_\Ks) = H(\XmS|\Wmi_\Ks)-H(\XmS|\Wmi_\Ks,\Zm)$.  Since user $k$ sends one of $M_{k0}M_{kx}$ codewords for each message, from \eqref{eqn:achR0} we have
\begin{align}
H(\XmS|\Wmi_\Ks) &= \log \paren{\textstyle \prod_{k=1}^K M_{k0}M_{kx}}\\
	\label{eqn:achprf4a}&= n \CW_\Ks
\end{align}

We can also write
\begin{equation}
\label{eqn:achprf4b}
H(\XmS|\Wmi_\Ks,\Zm) \le n\eta'_n
\end{equation}
where $\eta'_n \tozero$ as $n \toinf$ since, with high probability, the eavesdropper can decode $\XmS$ given $\Wmi_\Ks$ due to \eqref{eqn:achR0}.  Using \eqref{eqn:achRs}, \eqref{eqn:achR0}, \eqref{eqn:achprf3}, \eqref{eqn:achprf4a} and \eqref{eqn:achprf4b} in \eqref{eqn:achprf1}, we get
\begin{align}
\Deltac_\Ks 
	&\ge 1-\frac{\CW_\Ks-\CW_\Ks+\eta'_n}{\CM_\Ks-\CW_\Ks}
	\rightarrow 1\;\text{as }\eta'_n \tozero \label{eqn:G1}
\end{align}
The proof is completed by noting that due to \eqref{eqn:colsec}, $\Deltac_\Ks =1$ implies $\Deltac_\Ss =1, \, \forall \Ss \subseteq \Ks$, and writing
\begin{align}
\label{eqn:achprf0}
\Delta_\Ss \ge \frac{H(\Wmi_\Ss|\Zm)}{H(\Wm_\Ss)} 
	= \frac{H(\Wmi_\Ss)}{H(\Wm_\Ss)} 
	= \frac{\ssum_{k \in \Ss} \mu_k R_k}{\ssum_{k \in \Ss} R_k}
	\ge \delta
\end{align}
\end{IEEEproof}
We note that this way the achievable \dssr is
\begin{equation}
\label{eqn:RC}
\RC = \Min{\CM_\Ks, \dinv \bracket{\CM_\Ks - \CW_\Ks}}
\end{equation}

\subsection{Time-Division}
We can also use TDMA to get an achievable region.  Since, in such a scheme, only one user is transmitting at a given time, both sets of constraints collapse down to a set of single-user secrecy constraints, for which the results were given in \cite{leung-hellman:gaussianwiretap}:
\begin{theorem}
\label{thm:GT}
Consider this scheme: Let $\alpha_k \in [0,1], \, k=1,\dotsc,K$ and $\sum_{k=1}^K \alpha_k = 1$.  User $k$ only transmits $\alpha_k$ of the time with power $\Pmax_k/\alpha_k$ using the scheme described in \cite{leung-hellman:gaussianwiretap}.  Then, the following set of rates is achievable:
\begin{equation}
\label{eqn:RTalpha}
\bigcup_{\substack{\v{0} \preceq \vg{\alpha} \preceq \v{1} \\ \sum_{k=1}^K \alpha_k=1}} 
\braces{\Rm \colon R_k\le \min \braces{\frac{\alpha_k}{2\delta} \log \paren{\frac{\alpha_k+\Pmax_k}{\alpha_k+h \Pmax_k}},
\frac{\alpha_k}{2} \log \paren{1+\frac{\Pmax_k}{\alpha_k}}}, \quad k=1,\dotsc,K }
\end{equation}
We will call the set of all $\Rm$ satisfying the above, $\GsdT$.
\end{theorem}

\begin{IEEEproof}
Follows directly from \cite[Theorem 1]{leung-hellman:gaussianwiretap}.
\end{IEEEproof}

Note that with this scheme, the achievable \dssr is given by
\begin{equation}
\label{eqn:timeshare}
\RT[\delta,\vg{\alpha}] = \sum_{k=1}^K 	\min 
	\braces{\frac{\alpha_k}{2\delta} \log\paren{\frac{\alpha_k+\Pmax_k}{\alpha_k+h\Pmax_k}},
	\frac{\alpha_k}{2} \log\paren{1+\frac{\Pmax_k}{\alpha_k}}}
\end{equation}

\begin{theorem}
\label{thm:TDMAmaxsumcap}
The above described TDMA scheme achieves a maximum sum-rate of 
\begin{equation}
\label{eqn:RT}
\RT = \Min{\CM_\Ks, \dinv \bracket{\CM_\Ks - \CW_\Ks}}
\end{equation}
using the optimum time-sharing parameters 
\begin{equation}
\label{eqn:optTDMAparam}
\alpha_k^*=\frac{\Pmax_k}{\sum_{j=1}^K \Pmax_j}
\end{equation}
\end{theorem}
\begin{IEEEproof}
Maximizing each term in \eqref{eqn:timeshare} over the time-sharing parameters $\{\alpha_k\}$, is a convex optimization problem over $\alpha_k$.  Taking the derivative of the Lagrangian with respect to $\alpha_k$ and equating it to zero gives \eqref{eqn:optTDMAparam}, which simultaneously minimizes both terms in the minimum.  Using this in \eqref{eqn:timeshare} yields \eqref{eqn:RT}.
\end{IEEEproof}

Since in this scheme only one user is transmitting at any given time, both individual and collective constraints are satisfied.  We see that for collective secrecy constraints, this region is a subset of $\GsdC$.  For individual secrecy constraints, this does not hold.  We can then, using time-sharing arguments, find a new achievable region for individual constraints that is the convex-closure of the union of the two regions.
\begin{theorem}
The following region is achievable for individual secrecy constraints:
\begin{equation}
\label{thm:GIU}
\GsdIU = \text{convex closure of }\paren{\GsdI \cup \GsdT}
\end{equation}
\end{theorem}
\section{$\delta$-Secrecy Sum Capacity}
\label{sec:out}
In this section, we present an upper bound on the \dssr, denoted \Csum, for both individual and collective constraints, and show that this bound corresponds to the \dssr achievable under both constraints, giving us the secrecy sum-capacity of GMAC-WT for individual and collective constraints.  We note that a sum-rate constraint on both individual and collective constraints can be obtained using the constraints for the set $\Ks$.  In this case, both sets of constraints collapse down to 
\begin{equation}
\label{eqn:DeltaK}
\Delta_\Ks \triangleq \frac{H(\Wm_\Ks|\Zm)}{H(\Wm_\Ks)} \ge \delta
\end{equation}
\begin{theorem}
\label{thm:outer}
For the GMAC-WT, the \dssc for both individual and collective secrecy constraints is given by
\begin{equation}
\label{eqn:outer}
\Csum = \Min{\CM_\Ks, \dinv \bracket{\CM_\Ks - \CW_\Ks}}
\end{equation}
\end{theorem}
\begin{IEEEproof}
We first show that the right-hand side of \eqref{eqn:outer} is an upper bound on the \dssr for both constraints. Observe that \eqref{eqn:outer} is equal to the secrecy sum-rate achievable in \eqref{eqn:RC} for collective constraints using superposition coding, and by TDMA in \eqref{eqn:RT}, which satisfies both collective and individual constraints.  Hence, we get the $\delta$-secrecy sum-capacity of the GMAC-WT for both individual and collective constraints.

The first term in the minimum of \eqref{eqn:outer} is due to the converse for the GMAC, since the intended receiver needs to be able to decode the transmitted messages.  To see the second term, assume $\delta>0$.  This is without loss of generality, since if $\delta=0$, we have no secrecy constraint and only the first term applies. We first note that from Fano's inequality, we have
\begin{equation}
\label{eqn:outprf1}
H(\Wm_\Ks|\Ym,\Zm) \le H(\Wm_\Ks|\Ym) \le n \eta_n
\end{equation}
where $\eta_n \tozero$ as $n \toinf$.  We then use the constraint in \eqref{eqn:DeltaK}:
\begin{align}
R_\Ks &= \ninv H(\Wm_\Ks) \\
	&\le \ndinv H(\Wm_\Ks|\Zm) \\
	&\le \ndinv \bracket{H(\Wm_\Ks|\Zm) + n\eta_n - H(\Wm_\Ks|\Ym,\Zm)} \label{eqn:outprf1b}\\
	&= \ndinv I(\Wm_\Ks;\Ym|\Zm) + \eta_n' \\
	&\le \ndinv I(\Xm_\Ks;\Ym|\Zm) + \eta_n' \label{eqn:outprf2}
\end{align}
where we used \eqref{eqn:outprf1} in \eqref{eqn:outprf1b}, and $\Markov{\Wm_\Ks}{\Xm_\Ks}{\Ym}\rightarrow{\Zm}$ in the last step.  We now adopt Lemma 10 in \cite{leung-hellman:gaussianwiretap} to upper bound the differences between the received signal entropies at the receiver and wire-tapper:
\begin{lemma}[Lemma 10 in \cite{leung-hellman:gaussianwiretap}]
\label{lem:conv1}
Let $\xi = \ninv \h(\Ym)$ where $\Ym,\Zm$ are as given in \eqref{eqn:YZstd}. Then,
\begin{equation}
\h(\Ym)-\h(\Zm) \le n\xi - n\phi(\xi) \triangleq \frac{n}{2} \log \bracket{2\pi e \paren{1-h+\frac{h 2^{2\xi}}{2\pi e}}}
\end{equation}
\end{lemma}
\begin{IEEEproof}
The proof follows using the entropy power inequality \cite{cover-thomas:IT}. Recall that we can write $\h(\Zm)=\h(\sqrt{h}\Ym+\NMW)$. Then, using the entropy power inequality, we have
\begin{equation}
2^{\frac{2}{n} \h(\Zm)} = 2^{\frac{2}{n} \h(\sqrt{h}\Ym+\NMW)} 
	\ge 2^{\frac{2}{n} [\h(\Ym)+n\log\sqrt{h}]} +2^{\frac{2}{n} \h(\NMW)}
\end{equation}
Now $\h(\Ym) = n\xi$ and $\h(\NMW) = \frac{n}{2} \log [ 2 \pi e(1-h)]$.  Hence,
\begin{equation}
2^{\frac{2}{n} \h(\Zm)} \ge h 2^{2\xi} + 2 \pi e (1-h)
\end{equation}
which, after taking the log, gives
\begin{align}
\h(\Zm) &\ge \frac{n}{2} \log\bracket{h 2^{2\xi} +2\pi e(1-h)} \\
	&=\frac{n}{2} \log \bracket{2\pi e\paren{1-h+\frac{h2^{2\xi}}{2 \pi e}  }}
\end{align}
subtracting from $\h(\Ym)=n\xi$ completes the proof of the lemma.
\end{IEEEproof}
\begin{corollary}
\label{cor:conv1}
\begin{equation}
\h(\Ym) - \h(\Zm) \le n\bracket{\CM_\Ks-\CW_\Ks}
\end{equation}
\end{corollary}
\begin{IEEEproof}
From the converse to the GMAC coding theorem, we can show that
\begin{equation}
\h(\Ym) \le \frac{n}{2} \log \paren{2 \pi e (1+P_\Ks)}
\end{equation}
Let $\h(\Ym) = n\xi$.  Then, $\xi \le \onehalf \log \paren{2 \pi e (1+P_\Ks)}$, and since $\phi(\xi)$ is a non-increasing function of $\xi$, we get $\phi(\xi) \ge \phi\paren{\onehalf \log\paren{2\pi e(1+P_\Ks)}}$.  Thus, 
\begin{align}
\h(\Ym) - \h(\Zm) 
	&\le \frac{n}{2} \log \paren{2\pi e (1+P_\Ks)} -
		\frac{n}{2} \log \bracket{2 \pi e \paren{1-h+h(1+P_\Ks)}} \\
	&=n \bracket{\CM_\Ks-\CW_\Ks}
\end{align}\end{IEEEproof} 

Now, we can use \eqref{eqn:outprf2} to write
\begin{align}
I(\Xm_\Ks;\Ym|\Zm) &= I(\Xm_\Ks;\Ym,\Zm) - I(\Xm_\Ks;\Zm) \\
	&= I(\Xm_\Ks;\Ym) + I(\Xm_\Ks;\Zm|\Ym) - I(\Xm_\Ks;\Zm)  \label{eqn:outprf3a}\\
	&= I(\Xm_\Ks;\Ym) - I(\Xm_\Ks;\Zm) \\
	&= \h(\Ym) - \h(\Ym|\Xm_\Ks) - \h(\Zm) + \h(\Zm|\Xm_\Ks) \\
	&= \sumton \bracket{ \h(Z_i|\Xm_{\Ks,i}) - \h(Y_i|\Xm_{\Ks,i})} 
		+	\bracket{\h(\Ym)-\h(\Zm)} \\
	&=\bracket{ \frac{n}{2} \log \paren{2\pi e} 
		- \frac{n}{2} \log \paren{2\pi e }} + \bracket{\h(\Ym)-\h(\Zm)} \\
	&= \h(\Ym) - \h(\Zm) \\
	&\label{eqn:outprf3} \le n\bracket{\CM_\Ks - \CW_\Ks}
\end{align}
where we used $\Markov{\Xm_\Ks}{\Ym}{\Zm}$ to get \eqref{eqn:outprf3a}, and applied Corollary \ref{cor:conv1} in the last step.  Using \eqref{eqn:outprf3} in \eqref{eqn:outprf2} completes the proof.
\end{IEEEproof}

To see when we the secrecy constraint is more constraining on the sum-rate than the decodability constraint, we can write \eqref{eqn:outer} also as
\begin{equation}
\label{eqn:sumcapdelta}
\Csum = \case{\onehalf\log\paren{1+P_\Ks}}
	{\delta \le 1-\frac{\log(1+hP_\Ks)}{\log(1+P_\Ks)}}
	{\frac{1}{2\delta}\log\paren{\frac{1+P_\Ks}{1+hP_\Ks}}}
	{\delta \ge 1-\frac{\log(1+hP_\Ks)}{\log(1+P_\Ks)}}
\end{equation}
\section{Numerical Results and Observations}
\label{sec:results}

It can be seen that if the wire-tapper's channel is much worse than that of the legitimate receiver, i.e., $h \tozero$, then $\Csum \rightarrow C(P_\Ks)$, we incur no loss in \dssc.  On the other hand, if the wire-tapper's signal is not much degraded, i.e., $h \toone$, then $\Csum \tozero$, and it is no longer possible to communicate with secrecy. 

\begin{figure}[p]
\centering
\includegraphics[width=\figsize,angle=0]{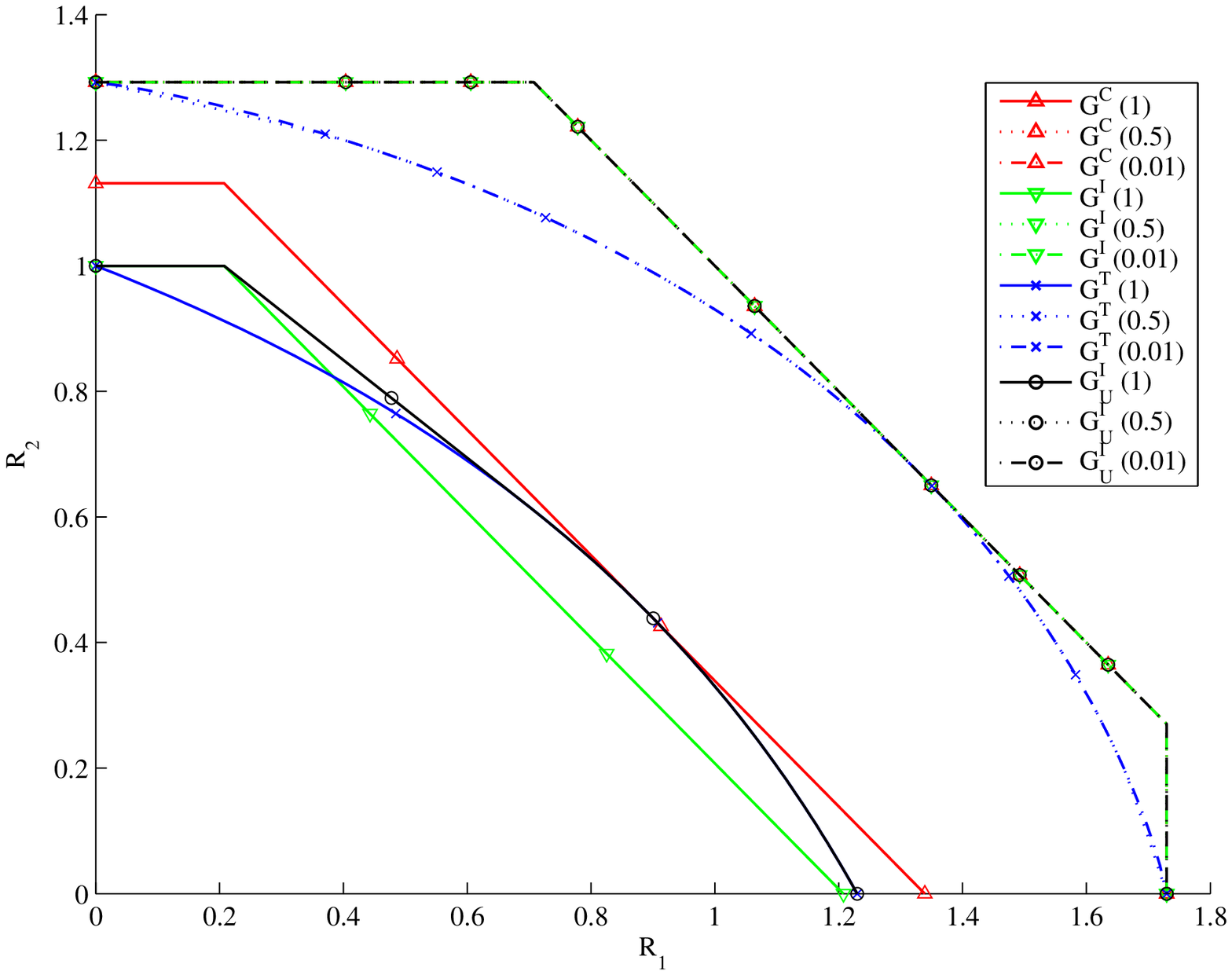}
\vspace{-0.1in}
\caption{Regions for $P_1=10$, $P_2=5$, $\delta=\{0.01, 0.5, 1\}$ and $h=0.1$}
\label{fig:C1}
\vspace{-0.15in}
\end{figure}

\begin{figure}[p]
\centering
\includegraphics[width=\figsize,angle=0]{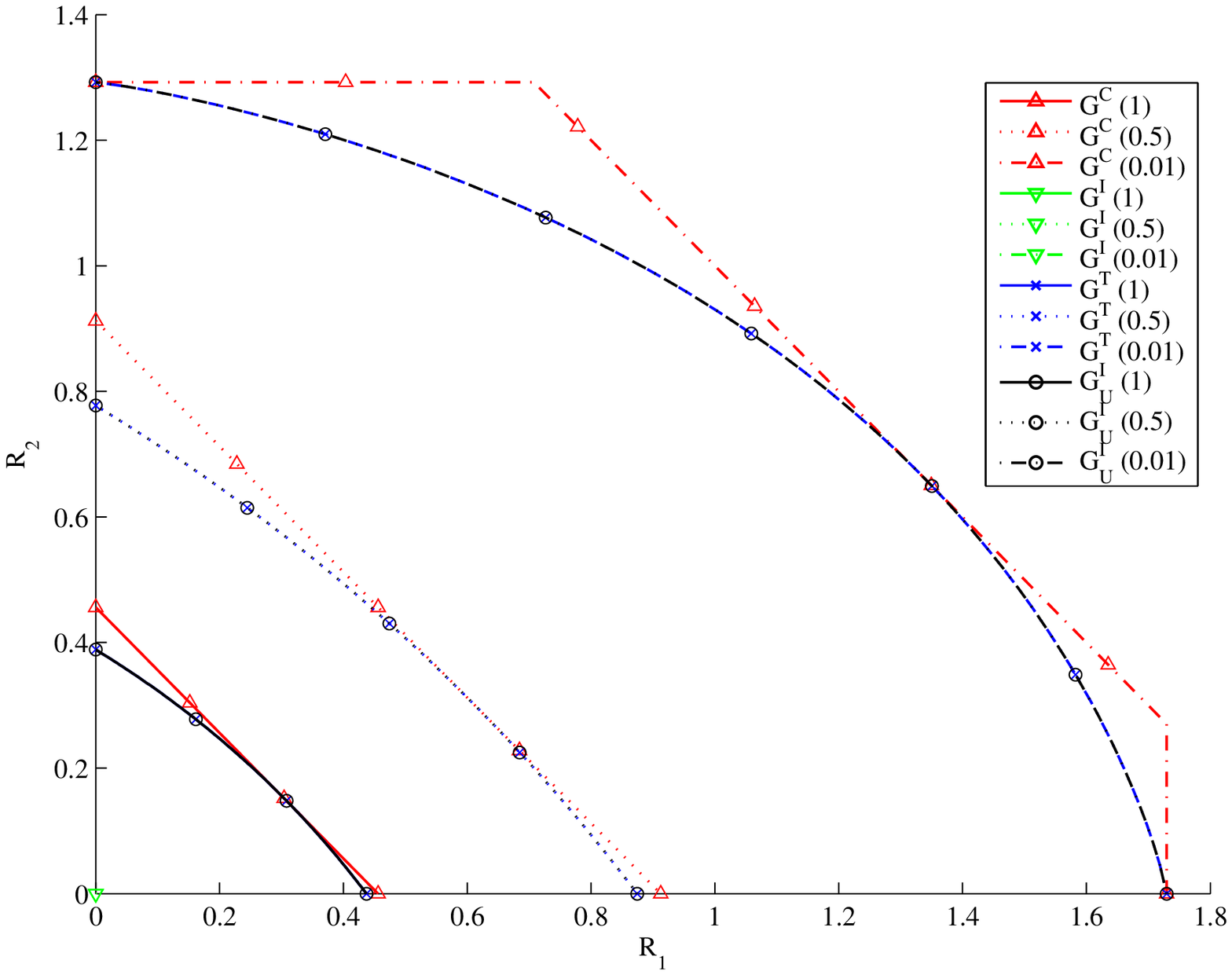}
\vspace{-0.1in}
\caption{Regions for $P_1=10$, $P_2=5$, $\delta=\{0.01, 0.5, 1\}$ and $h=0.5$}
\label{fig:C2}
\vspace{-0.15in}
\end{figure}

\begin{figure}[p]
\centering
\includegraphics[width=\figsize,angle=0]{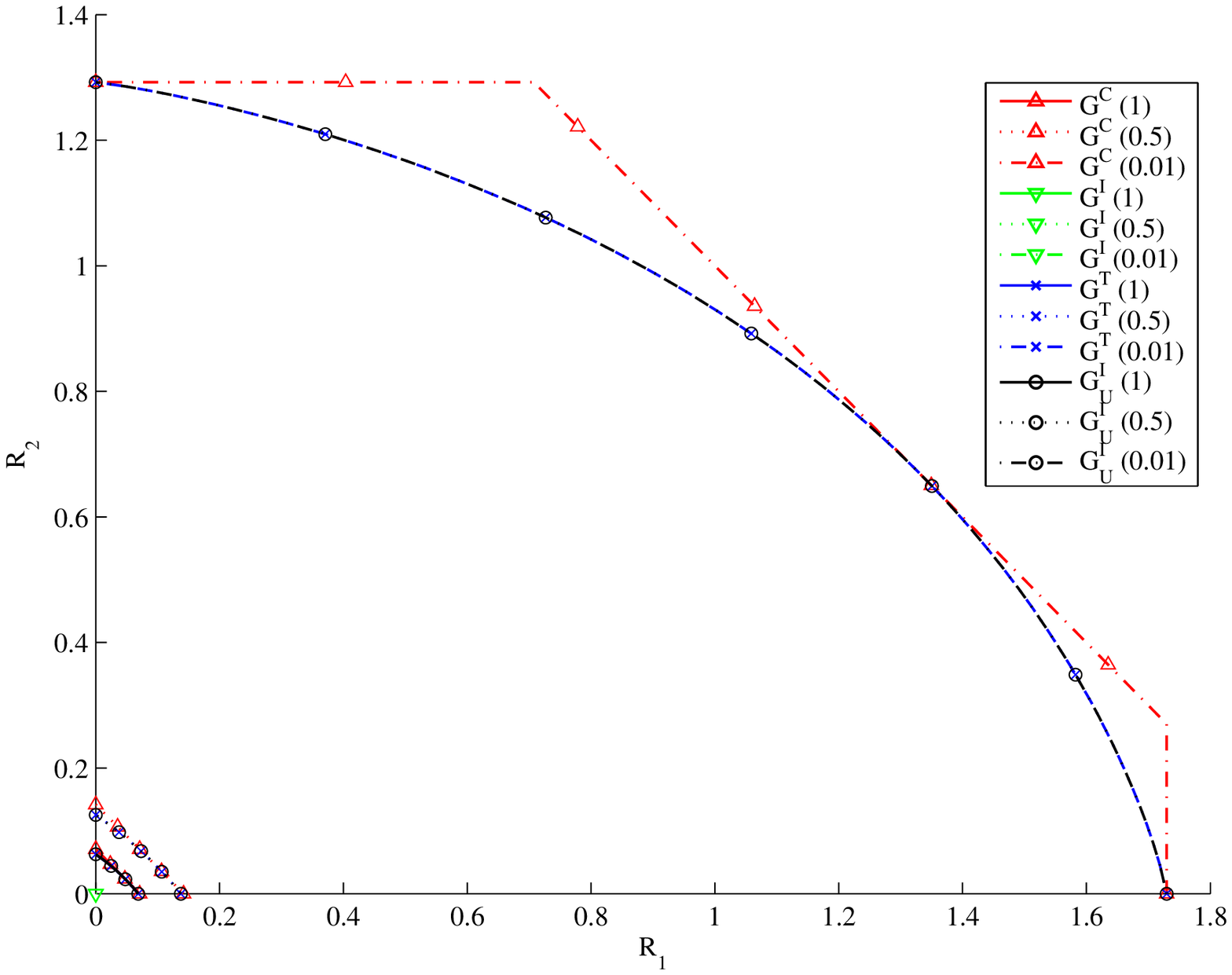}
\vspace{-0.1in}
\caption{Regions for $P_1=10$, $P_2=5$, $\delta=\{0.01, 0.5, 1\}$ and $h=0.9$}
\label{fig:C3}
\vspace{-0.15in}
\end{figure}

An interesting point to note is that the secrecy sum-capacity, $\Csum[\delta]$, is upper bounded by $\frac{1}{2\delta} \log\paren{\frac{1+P_\Ks}{1+hP_\Ks}}$, which is an increasing function of $P_\Ks$ for $h<1$, but as $P_\Ks \toinf$, $\Csum$ is upper bounded by $-\frac{1}{2\delta} \log h$.  We see that regardless of how much power we have available, the \dssc with a non-zero level of secrecy is limited by the channel's degradedness, $h$, and the level of secrecy required, $\delta$.  Also, it is inversely proportional to the level of secrecy desired, $\delta$, but inversely proportional to the logarithm of $h$, the degradedness of the channel.  Since in the range $[0,1]$, $\log(x)$ goes to $0$ faster than $-x^{-1}$, an increase in $h$ affects \dssc more than a similar increase in $\delta$.  This can be seen in Figures \ref{fig:C1}--\ref{fig:C3} which show the region $\Gs$ for $\delta=0.01,0.5,1$ and $h=0.1,0.5,0.9$ for two users.  When $\delta\rightarrow0$, we are not concerned with secrecy, and the resulting region corresponds to the standard GMAC region, \cite{cover-thomas:IT}.  The region for $\delta=1$ corresponds to the \ital{perfect secrecy} region, i.e., transmitting at rates within this region, it is possible to limit the rate of information leakage to the wire-tapper to arbitrarily small values.  It is seen that relaxing the secrecy constraint may provide a larger region, the limit of which is the GMAC region.  Note that it is possible to send at capacity of the GMAC and still provide a non-zero level of secrecy, the minimum value of which depends on the level of degradedness, $h$.  Especially when the degradedness is high, i.e., $h \tozero$, then we note that the achievable secrecy regions for $\delta=0.01$ and $\delta=0.5$ coincide with the GMAC region without the secrecy constraint.  Also shown in the figures are the regions achievable by the TDMA scheme described in the previous section.  Although TDMA achieves the secrecy sum capacity with optimum time-sharing parameters, this region is in general contained within $\GsdC$.  Depending on $h$ and $\delta$, the TDMA region is sometimes a superset of $\GsdI$, as observed in Figures \ref{fig:C1}, \ref{fig:C2}, sometimes a subset of $\GsdI$, as observed in Figure \ref{fig:C3} when $\delta=0.01$ or $\delta=0.5$, and sometimes the two regions can be used with time-sharing to enlarge the achievable region with individual constraints, see Figure \ref{fig:C3}, $\delta=1$. Close examination of these figures show that when the eavesdropper has a much worse channel, i.e., low $h$, and the secrecy constraint $\delta$ is low, then $\GsdI$ gives a larger region.  However, as we increase the secrecy constraint and the eavesdropper has a less noisy version of the intended receiver's signal, the TDMA region becomes more dominant.

Another interesting note is that even when a user does not have any information to send, it can still generate and send random codewords to confuse the eavesdropper and help other users when considering the collective secrecy constraints\footnote{This general idea is explored in detail in our follow-up work \cite{tekin:ALLERTON06, tekin:IT07a}.}.  This can be seen in Figures \ref{fig:C1}--\ref{fig:C3} as the TDMA region does not end at the ``legs" of $\GsdC$ when $\GsdC$ is not equal to the GMAC capacity region.  In addition, as noted in \cite{hellman-carleial:wiretap}, the intended receiver decodes the codeword transmitted completely, and as such $\floor{\dinv}$ low-rate messages can be transmitted each in perfect secrecy by the users.

\section{Conclusions}
We have examined secure communications in a multiple access channel in the presence of a wire-tapper.  Defining the appropriate secrecy measures, we have found achievable secrecy rate regions, and established the secrecy sum capacity of the GMAC-WT.

A main contribution of this paper is that, we show that the multiple-access nature of the channel can be utilized to improve the secrecy of the system.  Allowing confidence in the secrecy of all users, the secrecy rate of a user is improved since the undecoded messages of any set of users acts as additional noise at the wire-tapper and precludes it from decoding the remaining set of users.

The results in this paper are based on the wire-tapper having access to a degraded version of the intended receiver's signal.  The case where the eavesdropper's received signal is not necessarily degraded, termed the General Gaussian multiple access wire-tap channel, where some users may help improve the secrecy rates for other users by jamming the eavesdropper is explored in a follow-up work, \cite{tekin:IT07a}. The secrecy constraints in this paper are assumed to be identical across the users.  It might be interesting to explore heterogeneous scenarios where users might have different secrecy requirements.  We assume in this paper that the eavesdropper's channel gains are known to the legitimate parties.  These channel gains may not be easy to obtain in practice.  If the eavesdropper is known to be outside a certain area, we might opt to have a worst case system design, considering the boundary of the ``secure" area.

Finally, we remark that information theoretic secrecy has attracted a lot of attention in the research community including various multiterminal formulations since the submission of this work, for which we refer the reader to \cite{tekin:IT07a} and the references therein.

\bibliographystyle{IEEEtran}
\bibliography{IEEEabrv_mod,etekin_full}

\end{document}